\documentclass[12pt]{article}

\usepackage{newtxtext,newtxmath}

\usepackage{graphicx}

\usepackage[letterpaper,margin=1in]{geometry}

\renewenvironment{abstract}
	{\quotation}
	{\endquotation}

\date{}

\makeatletter
\renewcommand{\fnum@figure}{\textbf{Figure \thefigure}}
\renewcommand{\fnum@table}{\textbf{Table \thetable}}
\makeatother

\usepackage{scicite}

\usepackage{url}

\def\scititle{
	Dispersion-engineered meta-coverslip for 
    
    multimodal synthetic imaging
}
\title{\bfseries \boldmath \scititle}

\author{
	Yuchen Ma$^{1,2}$,
	Danlin Xu$^{1}$,
	Xinwei Wang$^{2}$,
	Guangwei Hu$^{2\ast}$,
	Liangcai Cao$^{1\ast}$\and
	\small\parbox{\textwidth}{\centering
	$^{1}$State Key Laboratory of Precision Measurement Technology and Instruments,
	Department of Precision Instrument, Tsinghua University; Beijing, China.
	}\and
	\small$^{2}$School of Electrical and Electronic Engineering, Nanyang Technological University; Singapore.\and
	\small$^\ast$Corresponding authors. Email: guangwei.hu@ntu.edu.sg (G.H.); clc@tsinghua.edu.cn (L.C.)\and
}

\begin{document} 

\maketitle

\begin{abstract} \bfseries \boldmath
Multimodal optical imaging provides comprehensive sample characterization but has traditionally been hindered by complex and bulky instrumentation. Here, we propose a dispersion-engineered meta-coverslip to seamlessly integrate bright-field, differential, fluorescence, and holographic imaging modalities within a standard microscope architecture, requiring no hardware modification or realignment. The meta-coverslip utilizes a scalable subwavelength multilayer film to engineer spatio-temporal dispersion for a customized high-dimensional transfer function. As an example, we demonstrate flexible switching among the four imaging modalities by simply tuning the illumination wavelengths. Lastly, we demonstrate that the synthesis of multimodal images can provide spatially registered structural and molecular information for more comprehensive biological analysis. Our approach provides an accessible, scalable, and flexible solution for advanced imaging and is extensible to other multiplexed optical systems in sensing and computing.
\end{abstract}

\newpage

\section{Introduction}

Multimodal optical imaging combines distinct imaging techniques to provide comprehensive sample characterization and has become a foundational tool across biological research, materials science, and industrial inspection~\cite{marti2010multimodality,stranks2021multimodal,bischof2024multimodal,songoptical}. Each imaging modality demands a specific optical configuration to accommodate the corresponding transfer function. For instance, bright-field, the most direct imaging modality, requires a uniform transfer function to preserve high spatial resolution. Phase contrast imaging enhances the visibility of transparent samples by adding a phase shift to zero-frequency components with customized masks. Holographic imaging typically utilizes interferometric setups to encode amplitude and phase via frequency-shifted interference patterns, allowing for quantitative characterization of sample morphology. While integrating diverse modalities into a single microscope greatly facilitates \textit{in situ} and spatially registered characterization by providing complementary information and hence has been explored extensively, the reliance on multiple optical paths and elements for functionality multiplexing inevitably results in cumbersome systems and complex alignment~\cite{dong2020super,kumar2020digital,chen2023artificial,sun2023fluorescence,kuppers2023confocal}.

Nanophotonics offers compact solutions for precise and completely decoupled light manipulation, enabling tailorable multifunctional optical elements. For instance, local metasurfaces have achieved success in the miniaturization of multidimensional sensing systems by engineering the meta-atom’s optical response to high-dimensional optical fields, including phase, spectral, polarization, and depth information~\cite{ye2023ultracompact,li2024single,kwon2020single,wang2023single,shen2023monocular,che2026varifocal,rubin2019matrix,zaidi2024metasurface,ji2025multidimensional,fu2025miniaturized,chen2026optical}. In a different approach, nonlocal flat optics modulate light directly in the wavevector space rather than real space~\cite{silva2014performing,shen2014optical,guo2018photonic,davis2019metasurfaces,kuai2019label,guo2020squeeze,xue2021high,he2022perfect,zhang2022incoherent,shastri2023nonlocal,li2023single,ma2024optogpt,tang2024metasurface,yao2024nonlocal,fan2024dispersion,shao2024multifunctional,qin2025disorder,chen2025nonlocal,ciabattoni2026nonlocal}. This unique characteristic inherently allows for plug-and-play optical processing and transfer function engineering, and has thus been exploited in applications such as edge detection and phase contrast imaging~\cite{chazot2020luminescent,zhou2020flat,wesemann2021nanophotonics,kuai2021planar,ji2022quantitative,chamoli2025nonlocal,pearson2025inverse,man2025versatile}. Recent demonstrations switch between two imaging modalities using phase-changing materials~\cite{cotrufo2024reconfigurable,yang2025nonlocal} or polarization transformations~\cite{cotrufo2023polarization,sulejman2025metasurfaces}. In these implementations, the number of independently accessible transfer functions is limited by the demonstrated material states or polarization channels. Integrating a larger set of distinct transfer functions within one compact device therefore remains challenging.

\begin{figure}[htbp]
	\centering
	\includegraphics[width=\textwidth]{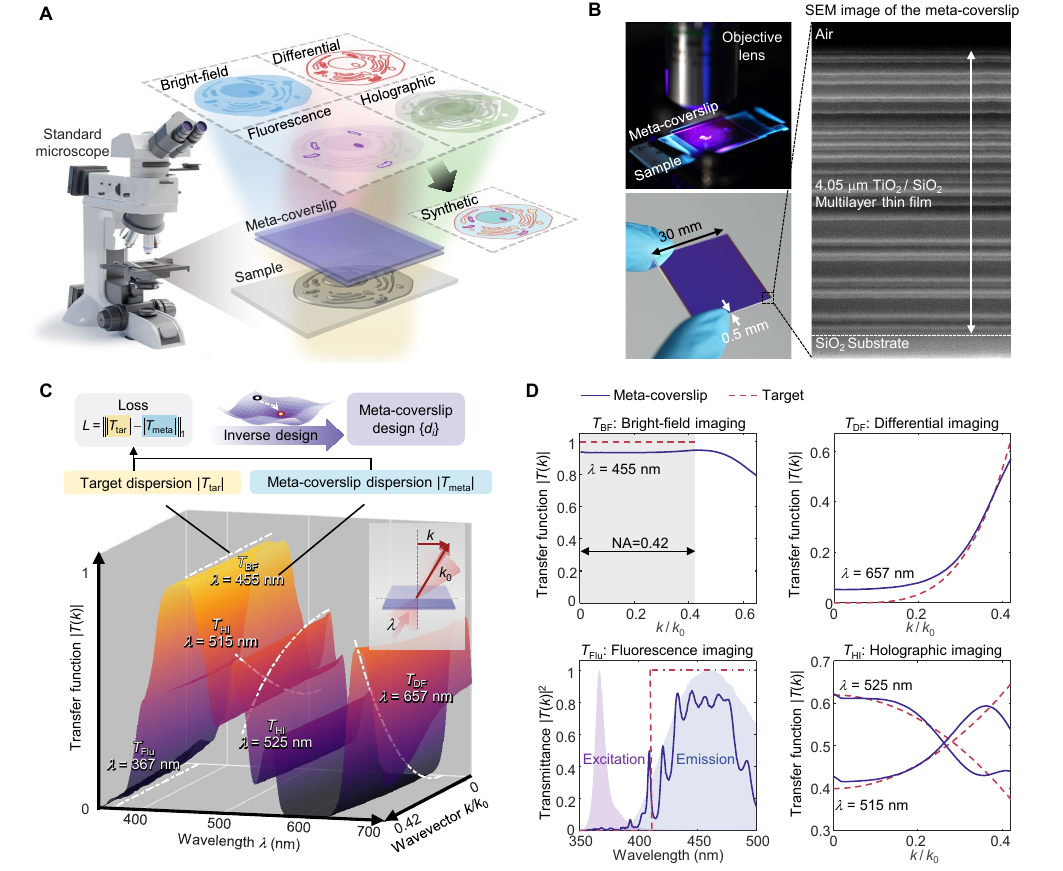} 
    
        \caption{\small
            \textbf{Concept of the dispersion-engineered meta-coverslip.} 
            (\textbf{A}) Schematic of the meta-coverslip-enabled multimodal imaging. A standard microscope is integrated with the meta-coverslip to provide wavelength-multiplexed access to four imaging modalities.
            (\textbf{B}) Photographs (left) and a scanning electron microscope (SEM) image (right) of the meta-coverslip, showing the compact, coverslip-like format and the TiO$_2$/SiO$_2$ multilayer structure.
            (\textbf{C}) The layer thicknesses are inverse-designed to minimize the loss between the simulated and target amplitude-transmittance profiles (dashed lines) over wavelength $\lambda$ and normalized transverse wavevector $k/k_0$.
            (\textbf{D}) Transfer function profiles tailored for the four modalities. Bright-field imaging uses an approximately uniform-transmission filter; differential imaging uses a high-pass filter that approximates spatial differentiation; fluorescence imaging uses a dichroic response to separate the excitation and emission bands; holographic imaging uses complementary high-pass and low-pass filters for phase encoding. NA, numerical aperture.
        }
\label{fig1} 
\end{figure}

Herein, we present a dispersion-engineered meta-coverslip that enables four distinct imaging modalities: bright-field (BF), differential (DF), fluorescence (Flu), and holographic imaging (HI). The meta-coverslip consists of a subwavelength multilayer thin-film stack with tailored dispersions in both spatial (wavevector, $k$) and temporal (wavelength, $\lambda$) domains, thereby featuring a high-dimensional optical transfer function, i.e., $T\left( k,\lambda  \right)$. Hence, various imaging modalities with corresponding transfer function as ${{T}_{i}}$ can be possibly supported within the subset ${{T}_{i}}\in T$, which is realized via resorting to the necessary wavelength channel ${{\lambda }_{i}}$ for our specific demonstration herein, as shown in Fig.~\ref{fig1}A. Working as a coverslip with a similar footprint, our flat optics device is readily compatible with a standard microscope architecture, allowing for seamless, plug-and-play integration without additional optical alignment (Fig.~\ref{fig1}B). Modality switching is achieved by simply varying the illumination wavelength, enabling \textit{in situ}, spatially registered multimodal imaging. We also demonstrate that the synthesized multimodal images yield higher information content, characterized by increased entropy and complementary features that are not fully captured by individual imaging modes. This plug-and-play solution not only enhances the accessibility and compactness of multimodal optical imaging but also demonstrates the potential of nanophotonic platforms for multiplexed optical modulation, paving the way toward high-dimensional sensing, imaging, and optical computing.

\section{Results}

\subsection{Design strategy of the meta-coverslip}

We begin by outlining the general principles of different imaging modalities and their associated optical designs with a unified physical picture through the optical transfer function engineering in both wavelength ($\lambda$) and spatial frequency ($k$) domains, i.e., $T\left( k,\lambda  \right)$. Unlike conventional 4$f$ systems that rely on lenses to transform light between real and spatial frequency domains, our meta-coverslip acts directly as a spatial frequency filter. Applying a specific transfer function $T\left( k,\lambda  \right)$ to the object $o\left( r,\lambda  \right)$ yields the modulated output~\cite{goodman2005introduction}:
\begin{equation}
   \tilde{o}\left( r,\lambda  \right)={{\mathcal{F}}^{-1}}\left\{ \mathcal{F}\left\{ o\left( r,\lambda  \right) \right\}\cdot T\left( k,\lambda  \right) \right\}, 
\end{equation}
where $\tilde{o}\left( r,\lambda  \right)$ represents the modulated light, $r=(x,y)$ is the real-space coordinate, and $\lambda$ is the wavelength selected for the corresponding modality. $\mathcal{F}$ and ${{\mathcal{F}}^{-1}}$ represent the Fourier transform and inverse Fourier transform, respectively. 

Each imaging modality corresponds to a unique transfer function configuration. As shown in Figs. 1C and 1D, bright-field imaging requires an all-pass transfer function, i.e., $|T\left( k,{{\lambda }_{\text{BF}}} \right)|=1$. Fluorescence imaging demands a wavelength-selective response characterized by low transmission at the excitation wavelength, i.e., $|T\left( k,{{\lambda }_{\text{Flu, excitation}}} \right)|^2 = 0$ and high transmission at the emission wavelength, i.e., $|T\left( k,{{\lambda }_{\text{Flu, emission}}} \right)|^2 = 1$. Differential imaging employs a high-pass filter $T\left( k,{{\lambda }_{\text{DF}}} \right)$, which converts phase variations into detectable intensity signals. Here, ${{\lambda }_{\text{BF}}}$ and ${{\lambda }_{\text{DF}}}$ denote the operating wavelengths for bright-field and differential imaging modalities, respectively; ${{\lambda }_{\text{Flu, excitation}}}$ and ${\lambda }_{\text{Flu, emission}}$ represent the excitation and emission wavelengths of the fluorophore, respectively. Holographic imaging reconstructs phase profiles by quantitatively modeling the phase variation-induced intensity signals. These transfer functions are not compatible with each other and hence cannot be supported on the same wavelength. This motivates our design of a nonlocal flat optical element with a customized high-dimensional transfer function in $(k,\lambda)$-domain such that each modality operates at a distinct and suitable wavelength. The much richer degree of freedom of wavelength compared with polarization or material states unlocks extended possibilities for multifunction integration.

To realize the desired transfer function $T(k, \lambda)$, we construct the meta-coverslip using a multilayer thin-film structure composed of alternating high- and low-refractive-index materials. Reflections and transmissions at layer interfaces introduce strong spatial and temporal dispersion, yielding wavelength- and angle-dependent transmittance. We compute $T(k, \lambda)$ using the transfer matrix method combined with a differentiable inverse-design procedure~\cite{born2013principles,ma2026sharpness}, which iteratively optimizes film thicknesses for target transmittances in the $(k, \lambda)$-domain at both $s$- and $p$-polarizations, as illustrated in Fig.~\ref{fig1}C. The designable layer thicknesses provide sufficient degrees of freedom to isolate the transfer functions required for different imaging modalities. Furthermore, this design is compatible with well-established, low-cost coating strategies for large-scale fabrication. Design and fabrication details are provided in Materials and Methods and Supplementary Text S1. 

\subsection{Characterization of the meta-coverslip}

Experimental validation confirms that the dispersion-engineered meta-coverslip achieves the designed transmittance profile. We characterize the intensity transmittance $|T(k,\lambda)|^2$ using a collimated broadband source, a linear polarizer, and a spectrometer while varying the incidence angle $\theta$. The transverse wavevector is $k=k_0\sin\theta$, where $k_0=2\mathrm{\pi}/\lambda$ is the free-space wavenumber. The measured intensity transmittance closely matches the simulation for both $s$- and $p$-polarizations, as detailed in Materials and Methods and Supplementary Text S2. Furthermore, the polarization-resolved 2D intensity-transmittance maps $|T(k_x,k_y,\lambda)|^2$ show isotropic features across the full NA. This weak polarization dependence supports consistent wide-field imaging performance across different azimuthal directions and polarizations. Additional details of the setup and measurement are provided in Materials and Methods and Supplementary Text S3.

Our meta-coverslip provides modality-specific transmission responses across different wavelength channels, as shown in Fig.~\ref{fig1}D. For BF imaging at 455 nm, it provides uniform transmittance, achieving an average amplitude transmittance of 0.896 across both polarizations within the objective's full NA of 0.42. For DF imaging, where the operating wavelength ${{\lambda }_{\text{DF}}}$ is 657 nm, the transmittance profile approximates $T\left( k,{{\lambda }_{\text{DF}}} \right)\propto {{({k}/{{{k}_{0}}})}^{4}}$, implementing fourth-order spatial differentiation to enhance high-frequency features. For Flu imaging using DAPI (4$^\prime$,6-diamidino-2-phenylindole) fluorophore with ultraviolet (UV) excitation at 367 nm and visible emission peaking at 460 nm, the meta-coverslip acts as a dichroic filter, blocking UV excitation (an average intensity transmittance of 0.006 at 367 nm) while transmitting the emission band (an average intensity transmittance of 0.776 over 445 to 475~nm). For HI imaging, the meta-coverslip provides high-pass and low-pass responses at 515~nm and 525~nm, respectively, enabling complementary spatial-frequency encoding for phase reconstruction.

The meta-coverslip also exhibits robustness to spectral broadening, maintaining the desired transfer functions over finite illumination bandwidths and facilitating operation under diverse illumination conditions (Supplementary Text S4). More broadly, the dispersion-engineering framework can be extended to larger NAs and broader wavelength ranges by increasing the accessible design space. Additional degrees of freedom, such as polarization, can further expand the available functional channels and enable more complex optical operations.

\subsection{Four-modality imaging}

\begin{figure}[htbp]
	\centering
	\includegraphics[width=\textwidth]{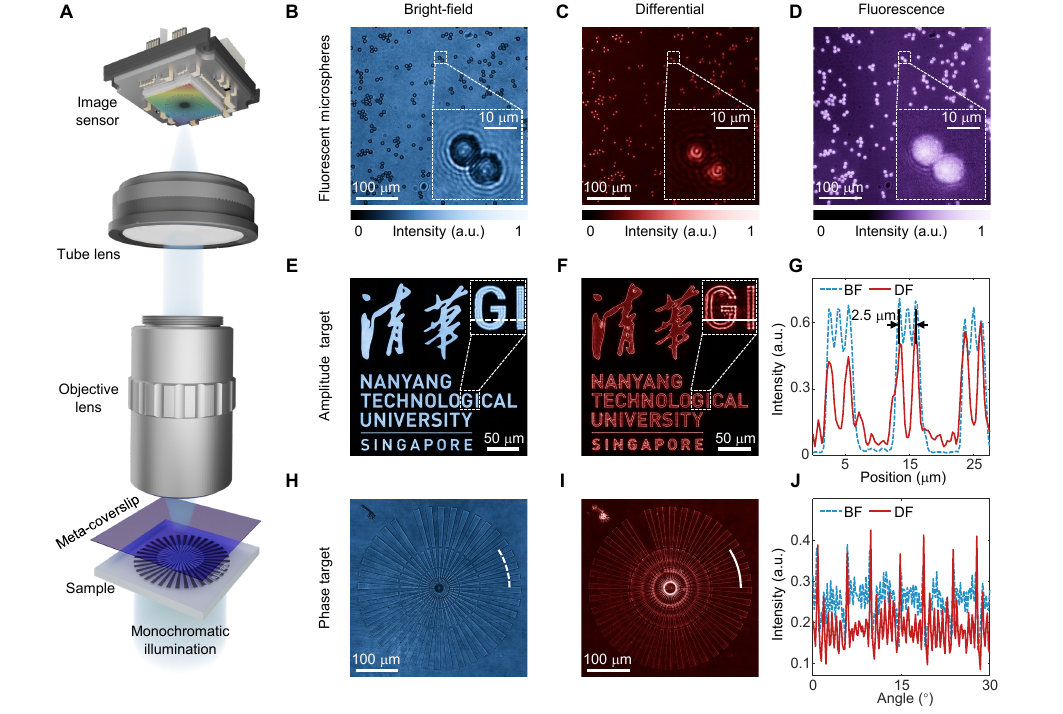} 
    \caption{\small
    \textbf{BF, DF, and Flu imaging with the meta-coverslip.}
    (\textbf{A}) Schematic of the experimental setup, which integrates the meta-coverslip into a standard upright-microscope architecture with collimated, wavelength-tunable illumination. The meta-coverslip is placed directly above the sample.
    (\textbf{B}-\textbf{D}) \textit{In situ} imaging of fluorescent microspheres using the BF, DF, and Flu modalities, respectively.
    (\textbf{E}-\textbf{F}) Imaging of a binary amplitude target using the BF and DF modalities, respectively. Insets show magnified views of the ``GI'' region.
    (\textbf{G}) Intensity profiles showing the enhanced edge contrast in DF imaging and resolved features with a linewidth of 2.5~\textmu m.
    (\textbf{H}-\textbf{I}) Imaging of a phase-only star target using the BF and DF modalities, respectively.
    (\textbf{J}) Intensity distributions extracted from the star pattern, showing the enhanced phase contrast provided by DF imaging.
    }
	\label{fig2} 
\end{figure}

To evaluate the multimodal imaging performance of our dispersion-engineered meta-coverslip, we conduct experiments using a standard upright microscope configuration shown in Fig.~\ref{fig2}A, equipped with a collimated monochromatic source with tunable wavelengths from 350 nm to 700 nm. The same objective lens with an NA of 0.42 is used throughout the experiments. The meta-coverslip is placed over the sample in a plug-and-play configuration without requiring optical setup modifications. Additional details of the microscopic imaging setup are provided in Materials and Methods and Supplementary Text S5.

We first validate wavelength-controlled switching among the bright-field, differential, and fluorescence modalities using 10-\textmu m-diameter polystyrene fluorescent microspheres (Fig.~\ref{fig2}B to D). Switching between imaging modalities is achieved by simply tuning the illumination wavelength, eliminating sample repositioning or objective realignment. In contrast, conventional implementations typically require modality-specific optical components, such as specialized prisms for differential interference contrast microscopy or dichroic filter cubes for fluorescence imaging. The wavelength-controlled switching therefore enables \textit{in situ} and spatially registered multimodal imaging with a consistent field of view, which is particularly advantageous for observing dynamic samples.

We characterize the BF and DF imaging performance using an amplitude mask and a phase-only target. In the BF modality, the meta-coverslip preserves the spatial resolution of the microscope, as further analyzed in Supplementary Text S5. In the DF modality, high-frequency components are selectively enhanced while the low-frequency background is suppressed, highlighting edge features of both amplitude and phase objects. For the binary amplitude mask (Figs. 2E to G), edge features with a linewidth of 2.5~\textmu m are clearly resolved with an average peak-to-dip contrast of 4.60. For the 250-nm-thick phase-only quartz target, DF imaging increases the signal-to-background contrast from 0.74 in BF to 1.52 (Figs. 2H-J), converting otherwise weak phase variations into visible edge contrast. The complementary capabilities preserve high-resolution bright-field imaging while adding differential contrast for both amplitude and phase features, benefiting comprehensive sample characterization.

\begin{figure}[htbp]
	\centering
	\includegraphics[width=\textwidth]{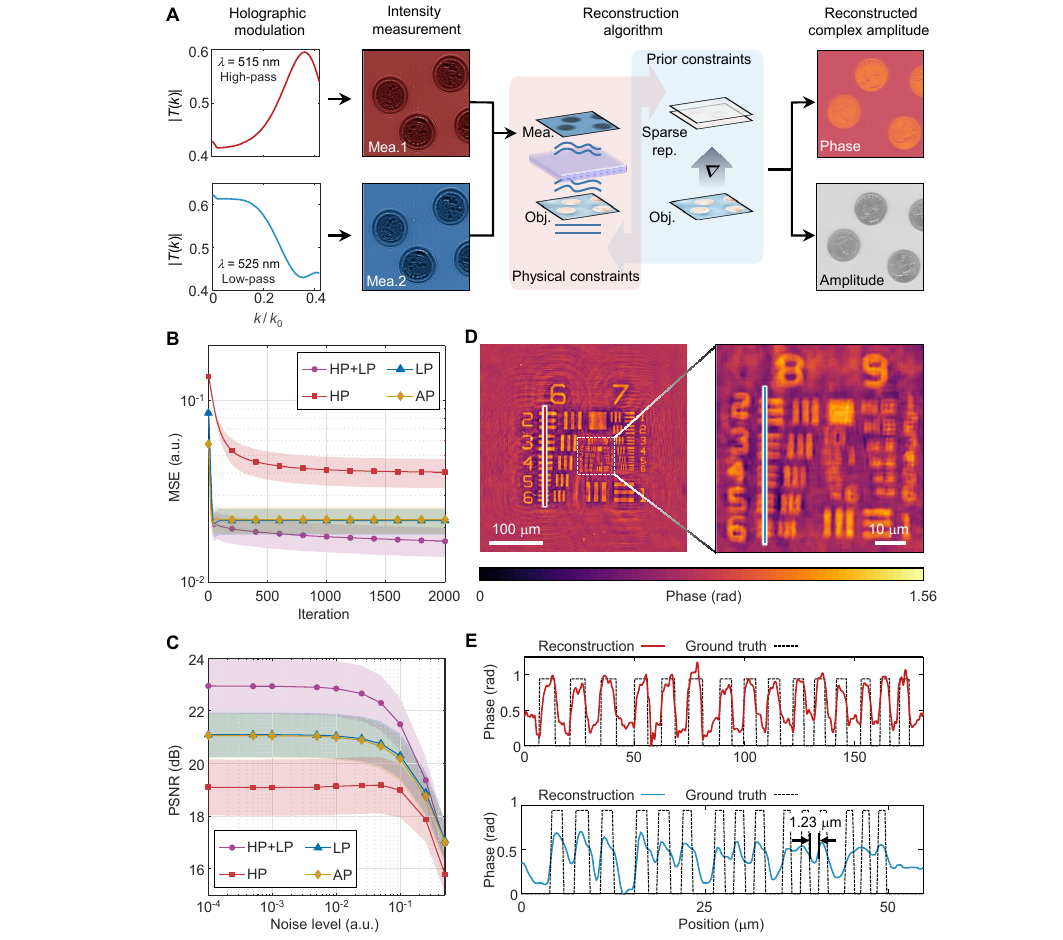} 
	\caption{\small
    \textbf{Holographic imaging with the meta-coverslip.}
    (\textbf{A}) Principle of holographic imaging with the meta-coverslip. Two intensity measurements are acquired at $\lambda=515$~nm and $\lambda=525$~nm, corresponding to high-pass (HP) and low-pass (LP) spatial-frequency filtering, respectively. The complex optical field is jointly recovered using a sparsity-regularized phase-retrieval algorithm.
    (\textbf{B}) Reconstruction convergence under four transfer-function configurations: HP, LP, all-pass (AP), and complementary HP+LP encoding.
    (\textbf{C}) Reconstruction fidelity under different noise levels for the four configurations. In (\textbf{B}) and (\textbf{C}), solid lines and shaded regions represent the mean and standard error of the mean, respectively.
    (\textbf{D}) Experimental quantitative phase reconstruction of a phase-only USAF 1951 resolution target.
    (\textbf{E}) Line profiles extracted from (\textbf{D}), comparing the reconstructed phase (solid lines) with the ground truth (dashed black lines) and demonstrating a half-pitch resolution down to 1.23~\textmu m.
    }
	\label{fig3} 
\end{figure}

Holographic imaging is implemented within a digital holography framework, with the meta-coverslip providing wavelength-dependent spatial-frequency modulation to assist phase encoding. Specifically, the meta-coverslip provides high-pass (HP) and low-pass (LP) spatial-frequency filtering at 515 nm and 525 nm, respectively (Fig.~\ref{fig3}A). With a slight defocus introduced for enhanced phase encoding, these two transfer functions modulate different spatial-frequency components of the object field and generate complementary intensity measurements. The complex field is then jointly recovered from the two measurements using a gradient-sparsity-regularized phase-retrieval algorithm (see Materials and Methods). This complementary spatial-frequency encoding provides additional measurement constraints for stable quantitative reconstruction.

We numerically evaluate the nontrivial contribution of the complementary HP and LP encoding by comparing it with HP-only, LP-only, and all-pass (AP) configurations. As shown by the reconstruction convergence in Fig.~\ref{fig3}B, HP-only encoding preserves high-frequency features but exhibits slower convergence due to greater sensitivity to noise, whereas LP-only encoding performs similarly to the AP configuration and provides limited additional constraints. In contrast, jointly using the HP and LP measurements substantially accelerates convergence and reduces the reconstruction error. The advantage is maintained over a broad range of noise levels as illustrated in Fig.~\ref{fig3}C, where the complementary encoding consistently achieves higher reconstruction fidelity than the individual filtering configurations. These results demonstrate that the two channels contribute complementary rather than redundant spatial-frequency information, improving both the well-posedness and noise robustness of quantitative phase reconstruction.

Experimentally, we validate the holographic modality using a phase-only USAF-1951 resolution target (Figs.~\ref{fig3}D and \ref{fig3}E). The reconstructed phase closely follows the ground-truth profile and resolves a half-pitch feature size down to 1.23~\textmu m, demonstrating both high spatial resolution and high-fidelity quantitative phase recovery. The meta-coverslip-enabled holographic modality therefore extends the system beyond intensity and contrast imaging to quantitative reconstruction of the complex optical field and supports structural analysis in biological imaging and defect inspection in industrial applications.

\subsection{Multimodal synthetic imaging}

\begin{figure}[htbp]
	\centering
	\includegraphics[width=\textwidth]{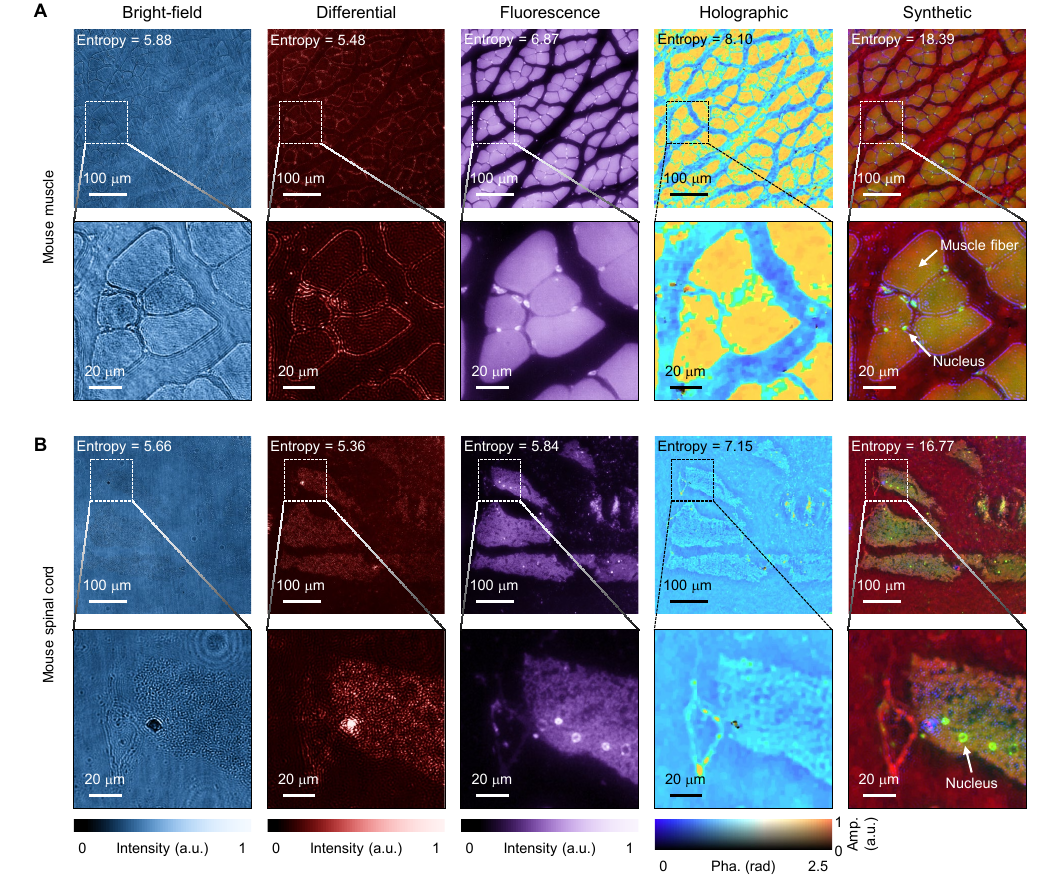} 
	\caption{\small
    \textbf{Multimodal imaging and synthesis of biological samples.} (\textbf{A}) Multimodal images of a DAPI-stained mouse muscle tissue section. From left to right: bright-field, differential, fluorescence, and holographic imaging. The rightmost panel is a synthetic false-color image, created by mapping the DF, Flu, and HI phase channels to blue, green, and red, respectively.
(\textbf{B}) Corresponding images for a DAPI-stained mouse spinal cord section, presented in the same order. For both samples, the lower panels show magnified views of the regions indicated by dashed boxes.
    }
	\label{fig4} 
\end{figure}

Finally, we demonstrate multimodal imaging of DAPI-stained mouse muscle and spinal cord tissue sections using the meta-coverslip (Figs.~\ref{fig4}A and \ref{fig4}B). The spatially registered modalities provide complementary biological information from the same tissue region. BF imaging captures the overall tissue morphology, while DF imaging enhances high-frequency features such as cellular and tissue boundaries. Fluorescence imaging specifically localizes DAPI-labeled nuclei, providing molecularly selective information. Holographic imaging maps optical-path variations associated with local tissue thickness and morphology, providing complementary quantitative structural information. Together, these modalities provide distinct views of cellular organization, molecular distribution, and tissue morphology that are difficult to obtain from any individual channel alone.

To directly visualize the relationships among these complementary features, we synthesize the DF, Flu, and HI phase data into a false-color image by mapping them to the blue, green, and red channels, respectively. In the muscle tissue, the fused image simultaneously reveals tissue boundaries, nuclear positions, and variations in tissue morphology. In the more structurally complex spinal cord tissue, it further allows the spatial relationship among nuclei, cellular boundaries, and surrounding tissue morphology to be directly identified within a single image. Such multimodal synthesis facilitates the correlation of structural and molecular features while avoiding registration errors between independently acquired modalities, providing a more comprehensive representation of the biological specimen.

We further quantify the complementary information provided by multimodal fusion using image entropy (see Materials and Methods). The synthesized images exhibit higher joint entropy than the individual modalities, with the joint entropy approaching the sum of the corresponding modal entropies, indicating limited statistical redundancy among the different imaging channels. This quantitative analysis supports the complementary nature of the structural, molecular, and morphological information revealed by multimodal imaging. By integrating these diverse information dimensions into a spatially registered representation, the approach provides a richer basis for comprehensive biological analysis and may further enable robust, data-driven diagnostics using multidimensional imaging data.

\section{Discussion}

We develop the dispersion-engineered meta-coverslip, a single passive optical component enabling seamless switching between four key imaging modalities: bright-field, differential, fluorescence, and holographic imaging. Unlike conventional systems requiring separate optical elements or multiple beam paths, our approach employs spatio-temporal dispersion engineering to unify these modalities within a single device. As a plug-and-play component suitable for scalable manufacturing, the meta-coverslip requires no modification to existing microscope hardware, lowering the barrier to advanced imaging adoption. The \textit{in situ} multimodal imaging capability eliminates mechanical realignment while preserving spatial registration, facilitating multimodal synthesis to provide comprehensive sample information.

While demonstrated here with a four-modality implementation, the underlying design framework is inherently scalable. By exploiting additional degrees of freedom in nanostructure design~\cite{park2022free,yin2024multi}, integrating active materials~\cite{gu2023reconfigurable,ha2024optoelectronic,li2024ultrafast,yoon2022miniaturized}, or advancing illumination control~\cite{zhou2025adaptive,gong2026spectral}, the approach may support additional imaging modalities and higher modality-switching speeds. Coupled with computational reconstruction, it can further enhance imaging quality and flexibility, opening opportunities in biological imaging, industrial inspection, and materials analysis. Beyond imaging, the same paradigm of high-dimensional transfer-function engineering can also be extended to applications such as optical manipulation~\cite{li2021integrating}, computing~\cite{hu2026metaoptics}, and lithography~\cite{kim2024nonlocal}.

In conclusion, the meta-coverslip exemplifies how nonlocal flat optics can address the longstanding trade-off between system complexity and functional diversity through high-dimensional transfer-function engineering. Beyond the demonstrated multimodal imaging system, this framework provides a scalable route toward next-generation optical platforms that combine miniaturized hardware with computational adaptability, enabling compact and information-rich systems for future photonic sensing and processing.

\section{Materials and Methods}

\subsection{Optimization and fabrication of the meta-coverslip}

The meta-coverslip consists of 35 alternating TiO$_2$ and SiO$_2$ layers with a total thickness of 4.05~\textmu m. The design optimization minimizes the mean absolute error between the target amplitude transmittance $|T_\text{tar}(k,\lambda)|$ and the simulated amplitude transmittance $|T_\text{meta}(k,\lambda)|$. The gradients of the loss with respect to the layer thicknesses are computed through automatic differentiation, and the layer thicknesses are iteratively updated using the Adam optimizer. The multilayer structure is fabricated using an ion-assisted deposition technique.

\subsection{Experimental dispersion characterization of the meta-coverslip}

The dispersion characteristics of the meta-coverslip are experimentally evaluated using an angle-resolved spectral-transmittance measurement system that measures the angle- and wavelength-dependent power transmittance. A broadband collimated source is generated by combining a visible light source (OSL2IR, Thorlabs) and a UV light source (LBD2000(25), LBTEC). After passing through a linear polarizer and the meta-coverslip mounted on a motorized rotation stage (GCD-011100M, Daheng), the transmitted light is analyzed by a spectrometer (HR6000, Ocean Optics).

\subsection{Multimodal microscopic imaging system}

The multimodal microscopic imaging system is constructed based on a standard upright microscope architecture, shown schematically and as a photograph in Fig.~\ref{S5_1}. For visible illumination, a supercontinuum laser (SuperK EVO HP EU-15, NKT Photonics) combined with a tunable filter (SuperK VARIA, NKT Photonics), pinhole, and achromatic lens provides narrowband illumination with a full width at half maximum of 10~nm. The UV illumination is generated by an LED (M365LP1, Thorlabs) and combined with the visible beam via a dichroic mirror (DMLP425R, Thorlabs) to ensure coaxial alignment. After passing through the sample and the meta-coverslip, the transmitted light is collected by an objective lens (NA = 0.42, 20$\times$ M Plan Apo, Mitutoyo), relayed by a tube lens (TTL 200, Thorlabs), and recorded by a CMOS camera (GT2050, Allied Vision).

\subsection{Holographic reconstruction algorithm}

For holographic imaging, the object is represented by a complex field $o(r)$, where $r$ denotes the real-space coordinate. Under illumination at wavelength $\lambda_i$, the meta-coverslip modulates the object field through its wavelength-dependent spatial-frequency transfer function $T(k,\lambda_i)$. Together with a slight defocus introduced into the imaging system, the forward model is written as
\begin{equation}
    m_i(r)
    =
    \left|
    \mathcal{M}_i\left\{o(r)\right\}
    \right|^2,
    \qquad
    \mathcal{M}_i\left\{o(r)\right\}
    =
    \mathcal{F}^{-1}
    \left\{
    \mathcal{F}\left[o(r)\right]
    T\left(k,\lambda_i\right)
    P\left(k,\lambda_i\right)
    \right\},
    \label{forward}
\end{equation}
where $m_i(r)$ is the measured intensity at wavelength $\lambda_i$, $\mathcal{M}_i$ denotes the corresponding forward operator, $k$ is the spatial frequency, and $\mathcal{F}$ and $\mathcal{F}^{-1}$ denote the Fourier and inverse Fourier transforms, respectively. $T(k,\lambda_i)$ represents the transfer function of the meta-coverslip, while $P(k,\lambda_i)$ is the phase-only free-space propagation transfer function associated with the introduced defocus. In this work, $\lambda_1=515$~nm and $\lambda_2=525$~nm correspond to the high-pass and low-pass transfer functions, respectively.

The complex object field is jointly reconstructed from the two intensity measurements by solving
\begin{equation}
    \hat{o}(r)
    =
    \underset{o(r)}{\arg\min}
    \sum_{i=1}^{2}
    \left\|
    \sqrt{m_i(r)}
    -
    \left|
    \mathcal{M}_i\left\{o(r)\right\}
    \right|
    \right\|_2^2
    +
    \beta
    \left\|
    \nabla o(r)
    \right\|_1,
    \label{min}
\end{equation}
where $\hat{o}(r)$ denotes the reconstructed complex field, $\nabla$ is the spatial-gradient operator, and $\beta$ is the regularization parameter. $\|\cdot\|_2$ and $\|\cdot\|_1$ denote the $\ell_2$ and $\ell_1$ norms, respectively. The first term enforces fidelity between the predicted and measured amplitudes, whereas the second term imposes gradient sparsity as a prior to regularize the reconstruction.

Equation~\ref{min} is solved by iteratively updating the complex object field using gradients of the data-fidelity and regularization terms. The $\ell_2$ data-fidelity term is differentiable, and its gradient is calculated analytically. The $\ell_1$ gradient-sparsity term is handled using its numerical subgradient, which is computed within the automatic-differentiation framework. Detailed derivations and implementation are provided in Supplementary Text S6. All experimental reconstructions use 2000 iterations.

For the numerical comparisons in Fig.~\ref{fig3}B and Fig.~\ref{fig3}C, the regularization weight $\beta$ is set to zero to isolate the contribution of the different transfer-function configurations. A total of 20 experimentally captured holographic samples of polystyrene beads and red blood cells are used as ground truth~\cite{lee2023deep}.

\subsection{Multimodal synthesis and entropy analysis}

The synthetic multimodal image is generated by mapping the DF, Flu, and HI phase channels to blue, green, and red, respectively. To enhance visualization in Fig.~\ref{fig4}, sigmoid functions are applied to the DF and Flu channels.

The image entropy is computed by treating each pixel as a multidimensional vector, depending on the number of image channels:
\begin{equation}
H = - \sum_{i=1}^N p_i \log_2 p_i,
\end{equation}
where $p_i$ is the probability of observing the $i$-th unique symbol formed by discretizing pixel values across all channels. For grayscale single-channel images (BF, DF, and Flu), pixel intensities are quantized into $B=256$ bins. For the HI modality, which generates complex-valued images (two channels: real and imaginary parts), both components are jointly quantized into $B^2$ bins. For RGB synthetic images (three channels), each channel is quantized into $B$ bins, resulting in up to $B^3$ unique symbols for the joint histogram. The joint histogram is then constructed, and entropy is calculated based on the resulting distribution.


\clearpage 

%
\bibliography{science_template.bib} 
\bibliographystyle{scienceadvances.bst}



\end{document}